\documentclass[letterpaper]{jpconf}
\usepackage{graphicx}

\newcommand{\dblfig}[7]{
  \begin{figure}[tbh]
    \centering
    \includegraphics[width=#2]{#1}%
    \includegraphics[width=#4]{#3}
    \caption[#7]{#6}
    \label{fig:#5}
  \end{figure}
}

\begin{document}
\title{Proton Structure Measurements and the HERAPDF fit}

\author{Jan Kretzschmar for the H1 and ZEUS Collaborations}

\address{University of Liverpool}

\begin{abstract}
New and previously published measurements on the deep inelastic $ep$
scattering cross section by the H1 and ZEUS collaborations are
presented. The uncertainties can be significantly
reduced by a model independent combination procedure, which treats the
systematic error correlations in a coherent way. The combined H1 and
ZEUS measurements of the inclusive neutral and charged current cross
sections are used to perform a common NLO QCD fit, called HERAPDF $0.1$. The
resulting set of parton density functions (PDFs) have a much improved
experimental uncertainty compared to previous extractions using the
uncombined H1 or ZEUS data.
\end{abstract}

\section{Introduction}

The HERA collider facility in Hamburg, Germany, was a unique machine for
lepton-proton scattering at highest energies. For the two experiments
H1 and ZEUS protons with an energy of up to $920$~GeV and electrons or
positrons with an energy of $27.6$~GeV were collided. This is
equivalent to a maximal centre of mass energy of $\sqrt{s} = 320$~GeV.
At the end of June 2007 the data taking finished.

In Deep Inelastic Scattering (DIS) of leptons off nucleons the
substructure of the nucleons was discovered. DIS continues to be the
tool for exploring the substructure of the nucleons with high
precision, i.e. measuring their quark and gluon content in the form of
so called parton distribution functions (PDFs). The evolution of PDFs
is a sensitive test of our understanding of QCD dynamics, which is
expressed in the form of PDF evolution equations. Furthermore a
precise knowledge of PDFs is vital for measurements at hadron
colliders, such as the LHC.

The kinematics of the scattering are described in terms of the Lorentz
invariant quantities: the Bjorken scaling variable $x$, the
inelasticity $y$, and the virtuality $Q^2$. The HERA experiments have
made measurements of the proton structure for $Q^2$ values of up to
$50000$\,GeV$^2$ and $x$ values down to $10^{-6}$.

One of the most fundamental measurements to be performed is that of the
neutral current (NC) inclusive cross section for the reaction $ep \rightarrow e'X$,
which can be expressed in the form
\begin{equation} \label{sigmar}
    \frac{\mathrm{d}^2\sigma^{e^{\mp}p}_{NC}}{\mathrm{d}x\mathrm{d}Q^2} 
    = \frac{2\pi\alpha^2Y_+}{xQ^4}
    \left( F_2
       - \frac{y^2}{Y_+}  F_L 
      \pm \frac{Y_-}{Y_+}  xF_3 
    \right)\;,
\end{equation}
with $Y_\pm = 1 \pm (1-y)^2$ and the structure functions $F_2$, $F_L$
and $xF_3$.
At leading order, the structure functions relate to the PDFs as
\begin{equation}
  F_2  =    x\sum e_q^2 (q(x) + \bar{q}(x) )\;\;\;\mathrm{and}\;\;\; xF_3 \sim x\sum 2 e_q a_q( q(x) - \bar{q}(x))\,.
\end{equation}

Additional information can be obtained from charged current (CC) cross
section for the reaction $ep \rightarrow \nu_eX$:
\begin{equation}
  \sigma^{CC}_{e^+p} \sim x(\bar{u} + \bar{c}) + x(1-y)^2(d + s)\;\;\;\mathrm{and}\;\;\;
  \sigma^{CC}_{e^-p}  \sim x( u + c ) + x(1-y)^2 (\bar{d} + \bar{s})\,.
\end{equation}

The gluon density $xg(x, Q^2)$, which does not enter the inclusive
cross section calculations at leading order, is constrained by the $F_2$
scaling violations, jet cross sections and the measurement of $F_L$.

\section{Combination of HERA cross section data and QCD Fit}

Both ZEUS and H1 have performed inclusive cross section measurements
with comparable precision and covering a similar kinematic range.
Therefore the results of the two experiments may be combined to
provide a single set containing all inclusive HERA cross sections and
to be used for further QCD analyzes. The averaging procedure,
described in detail in~\cite{h1lowq2}, is without theoretical
assumptions.
Only small theoretical corrections are needed for
swimming points to a common $x, Q^2$ grid and for adjusting earlier
measurements performed at a proton beam energy of $820$\, GeV.

The precision cross section data from the HERA experiments are
typically reported with three different components of the measurement
uncertainty: statistical and systematic uncertainties, where the
latter consist of parts uncorrelated or correlated between different
kinematic domains. The correlated uncertainties, which are due to
effects like the luminosity measurement or shifts in the electron
energy calibration need to be treated correctly in the averaging
procedure. As a result, not only the uncorrelated and statistical part
of the uncertainties are reduced. Because the measurements by the two
experiments have different sensitivities to the correlated
uncertainties, these can be constrained further and thus reduced.

\dblfig{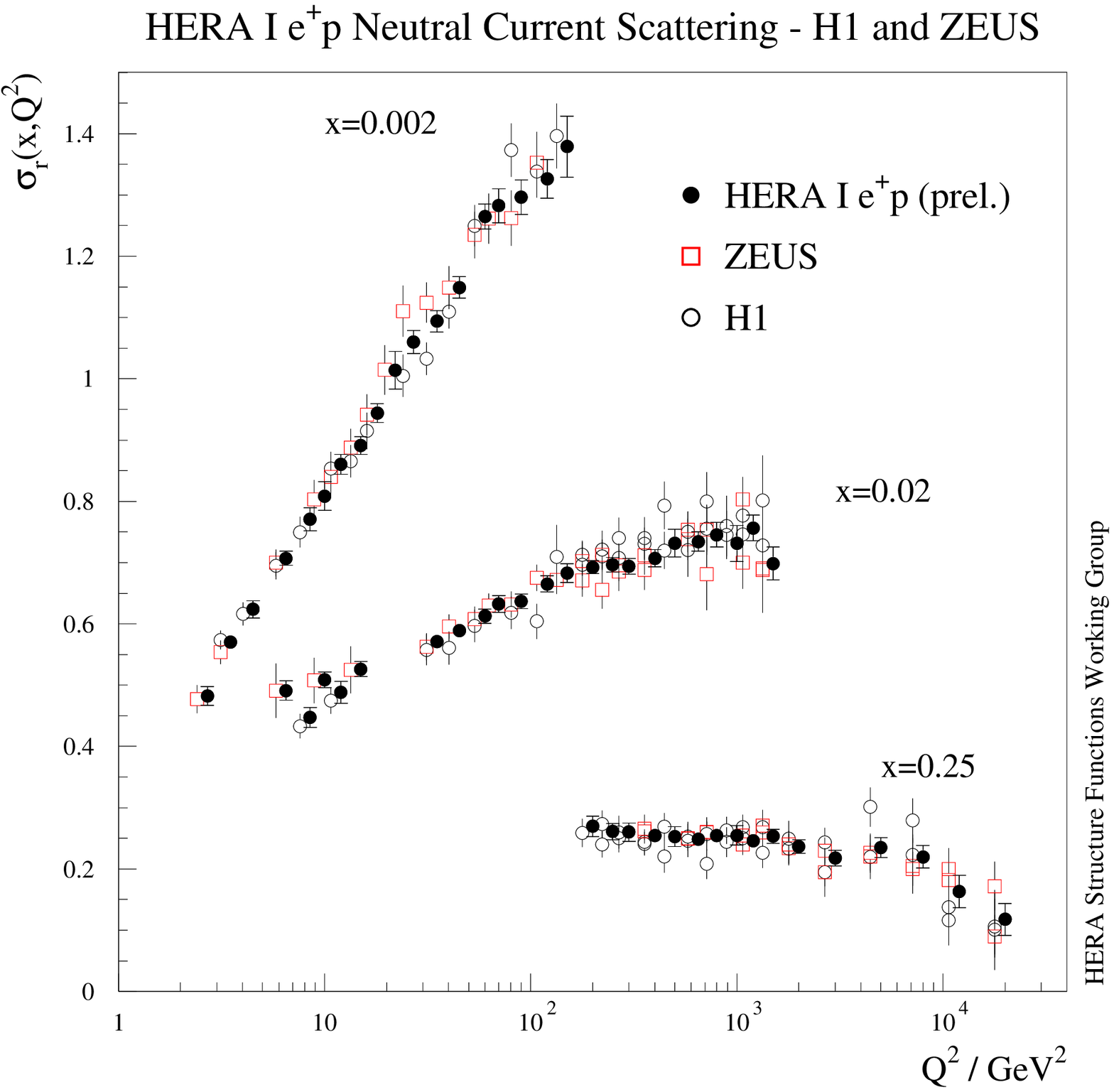}{0.5\linewidth}
{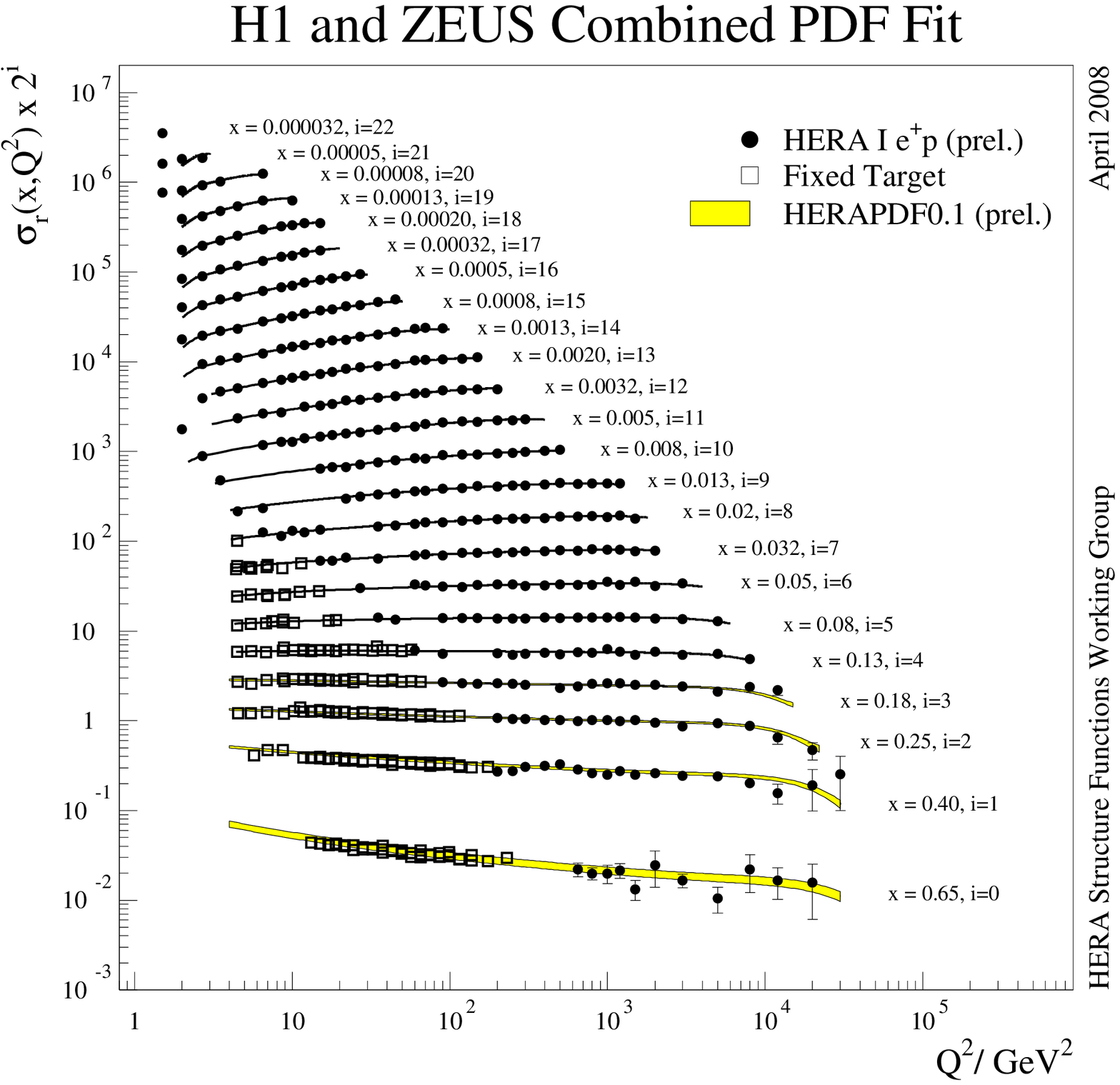}{0.5\linewidth}{h1zcombfit} {Left: Published
  H1 and ZEUS data and their combination for a set of fixed $x$ values
  as a function of $Q^2$. Right: The full combined $e^+p$ data set
  shown together with the QCD analysis result HERAPDF $0.1$~\cite{h1zeuscomb}. The HERA
  data is complemented by fixed target data at lower values of $Q^2$
  and higher $x$.}

The combination procedure is performed separately for the sets of
$e^+p$ and $e^-p$ scattering and using both NC and CC
data~\cite{h1zeuscomb}. The original data and the combination are
shown for a few exemplary values of $x$ of the $e^+p$ NC data set in
figure \ref{fig:h1zcombfit}, left. A remarkable reduction of the
uncertainties is apparent. Figure \ref{fig:h1zcombfit}, right, shows
the full $e^+p$ data set, which is seen to cover four orders of
magnitude in $x$ and $Q^2$ with very high precision. The compatibility
of the two experiments is observed to be very good with 
a total $\chi^2 = 510$ for $599$ averaged points.

The combined data is in the following used to perform a QCD analysis
at NLO using HERA data only. For the analysis, which is called HERAPDF
$0.1$, the DGLAP evolution equations are used to evolve the PDFs,
which are parametrized at the starting scale of $Q_0^2 = 4$\,GeV$^2$.
A full evaluation of experimental and theory model
uncertainties is performed. The resulting PDFs are shown in figure
\ref{fig:qcdfits} at a scale of $Q^2 = 10$\,GeV$^2$ and compared to
earlier separate analyzes of the H1 and ZEUS
collaborations~\cite{h1pdf2000, zeusjets}. A reduction of the
uncertainties is seen, which is due to the improved combined data used
as input. This is most notable on the gluon density, which dominates
all other partons at low $x$ whose uncertainty is typically also
larger than the quark PDFs, which are directly accessed in DIS.

\dblfig{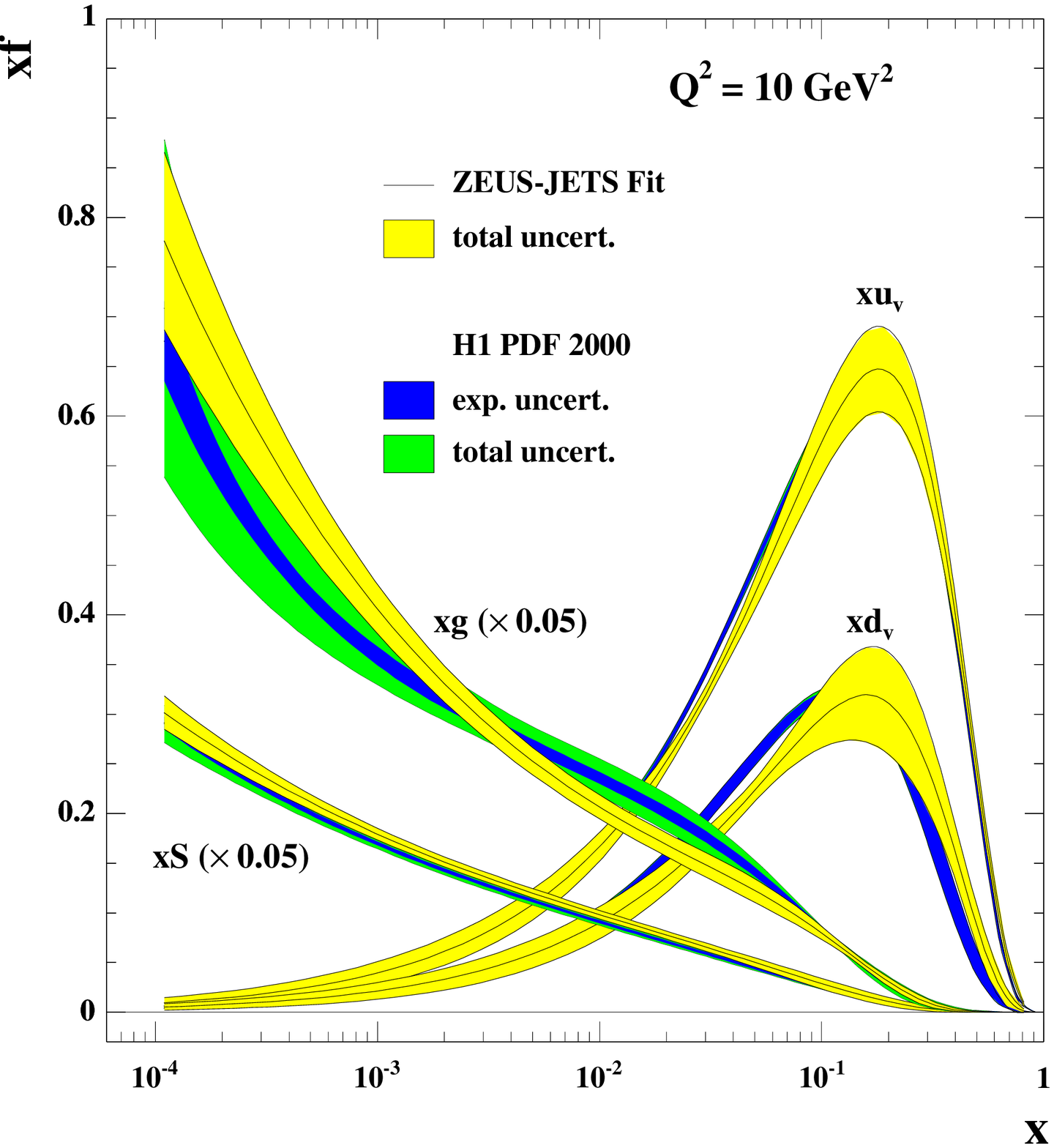}{0.435\linewidth}
{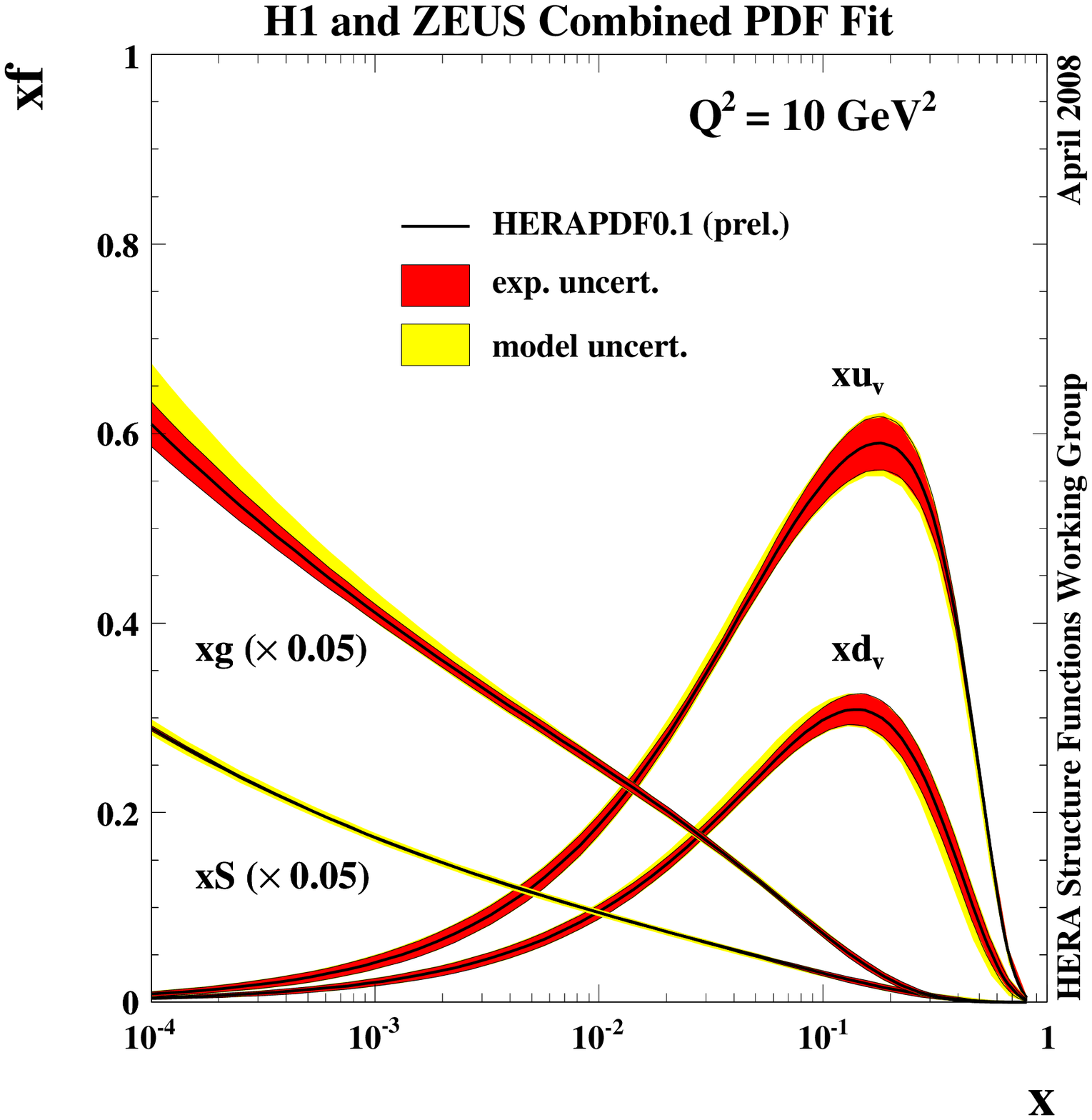}{0.5\linewidth}{qcdfits} {Published H1 and
  ZEUS QCD fits (left)~\cite{h1pdf2000, zeusjets} compared to the new
  analysis using the combined HERA data (right)~\cite{h1zeuscomb}. For
  all fits, the valence distributions, $xu_v$ and $xd_v$, the total
  sea quark distribution $xS$ and the gluon distribution $xg$ are
  shown at a scale of $Q^2 = 10$\,GeV$^2$. }

\section{New Inclusive Measurements from HERA}

While data taking at HERA has stopped, more precise measurements have
become available recently or are expected for the near future.

In general, analyzes at low and intermediate $Q^2 \leq 150$\,GeV$^2$
can be performed using relatively small data sets. Therefore the new
high precision analyzes of H1~\cite{h1lowq2,h1medq2} were performed
using data from the HERA-I running period, which ended in the year
2000. The total measurement uncertainties were reduced to as low
as $1.5\%$ per point, which is about a factor of two better than
previous data covering this kinematic domain~\cite{h197, zeus97}.
As an example, the new results on the structure function $F_2$
from~\cite{h1medq2} are shown in figure~\ref{fig:newdata}, left.

The analyzes at higher $Q^2 \geq 150$\,GeV$^2$ are in general more
constrained by the available statistics. Therefore improvements are
expected from the analysis of the HERA-II data, where nearly
$400$\,pb$^{-1}$ were collected per experiment
after the HERA luminosity upgrade starting from the year 2003. The
data sample is about balanced between $e^+p$ and $e^-p$. A new feature
of the HERA-II data is also beam polarization, where typical average
values of $30-40\%$ are reached.

ZEUS has recently published new results on the inclusive NC and CC
$e^-p$ cross section~\cite{zeusnewnc, zeusnewcc}.
Figure~\ref{fig:newdata}, right, shows the CC cross sections and
highlights their dependence on different quark flavours in the proton.
These are the first
double differential inclusive measurements using the polarized electron
beam. Also the employed $e^-p$ HERA-II data samples have an integrated
luminosity ten times larger than the previously used HERA-I sample.

\dblfig{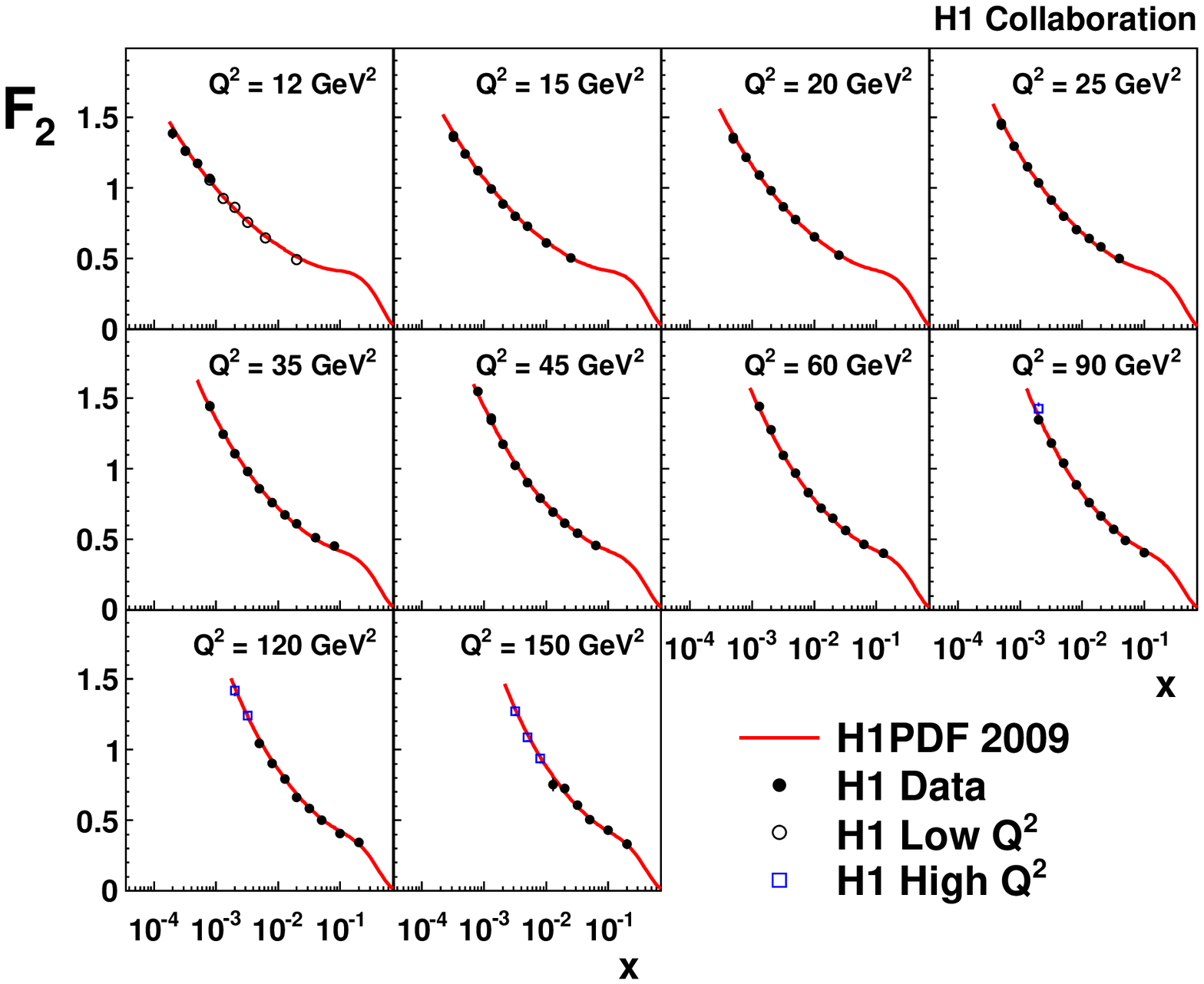}{0.56\linewidth}
{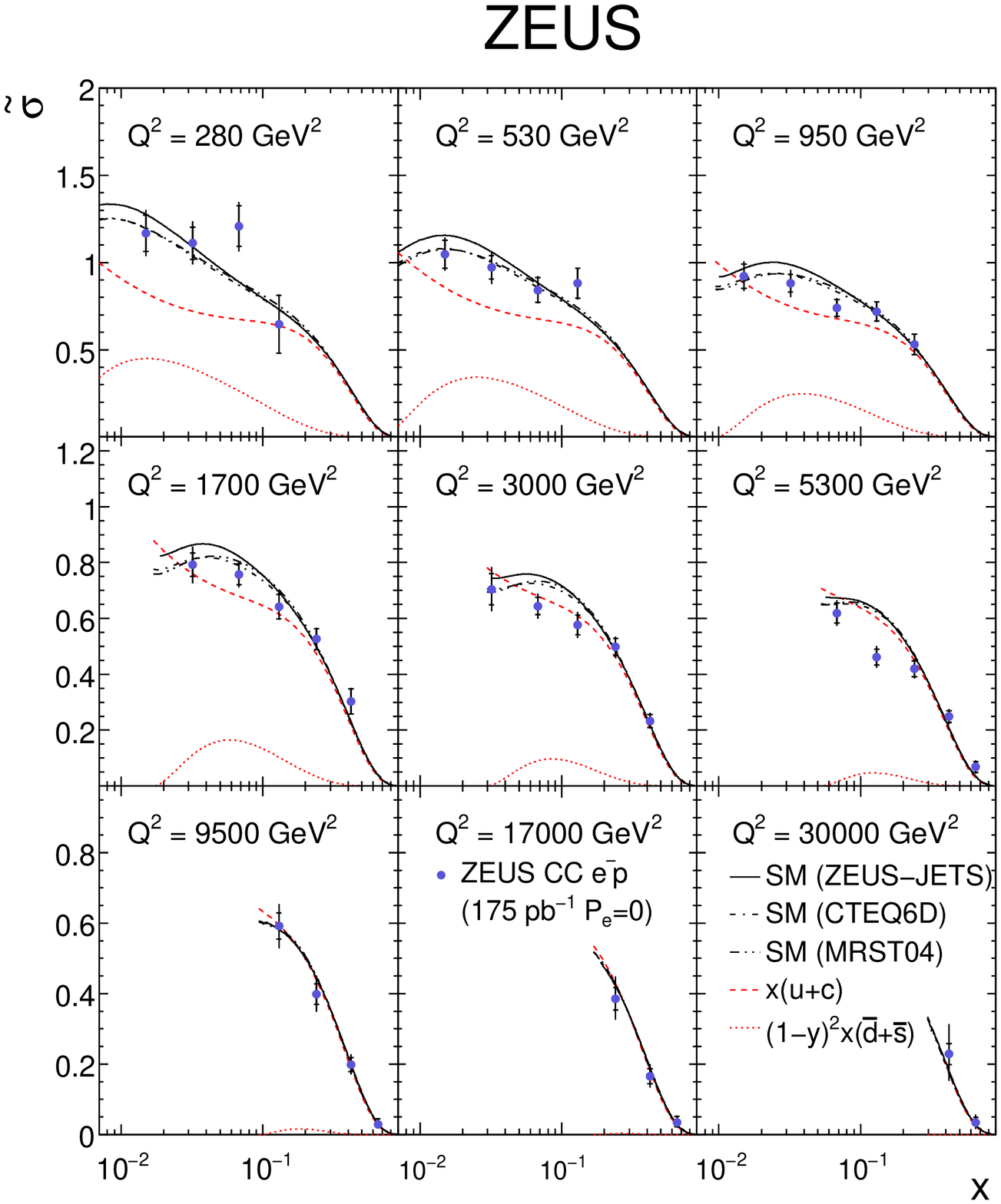}{0.43\linewidth}{newdata} {New results on the
  inclusive NC cross section at lower $Q^2$ by H1
  (left)~\cite{h1medq2} and on the inclusive CC cross section at lower $Q^2$ by ZEUS
  (right)~\cite{zeusnewcc}. The data are compared to different QCD fits.}

\section{Conclusions}

The HERA experiments measure the proton structure in a wide range of
$x$ and $Q^2$, thus providing stringent tests of QCD and valuable
input for the LHC. A new procedure combines the already published data
to obtain one HERA data set with improved uncertainties. A combined
QCD fit, HERAPDF 0.1, is performed using this combined data set. It is
able to describe the data and provides PDFs with much reduced
uncertainties compared to previous fits performed separately by the H1
and ZEUS collaborations. In addition new measurements are available
with significantly increased experimental precision.

\section{References}

\bibliographystyle{iopart-num}

\begin{thebibliography}{99}

\bibitem{h1lowq2}
  F.D. Aaron {\it et al.} [H1 Collaboration],
  arXiv:0904.0929 [hep-ex].

\bibitem{h1zeuscomb}
  H1 and ZEUS Collaborations,
   H1prelim-09-045, ZEUS-prel-09-011.

\bibitem{h1pdf2000}
  C.~Adloff {\it et al.}  [H1 Collaboration],
  Eur.\ Phys.\ J.\ C {\bf 30} (2003) 1
  [arXiv:hep-ex/0304003].

\bibitem{zeusjets}
  S.~Chekanov {\it et al.}  [ZEUS Collaboration],
  Eur.\ Phys.\ J.\ C {\bf 42} (2005) 1
  [arXiv:hep-ph/0503274].

\bibitem{h1medq2}
  F.D. Aaron {\it et al.} [H1 Collaboration],
  arXiv:0904.3513 [hep-ex].


\bibitem{h197}
  C.~Adloff {\it et al.}  [H1 Collaboration],
  Eur.\ Phys.\ J.\ C {\bf 21} (2001) 33
  [arXiv:hep-ex/0012053].

\bibitem{zeus97}
  S.~Chekanov {\it et al.}  [ZEUS Collaboration],
  Eur.\ Phys.\ J.\  C {\bf 21} (2001) 443
  [arXiv:hep-ex/0105090].

\bibitem{zeusnewnc}
  S.~Chekanov {\it et al.}  [ZEUS Collaboration],
  arXiv:0901.2385 [hep-ex].

\bibitem{zeusnewcc}
  S.~Chekanov {\it et al.}  [ZEUS Collaboration],
  arXiv:0812.4620 [hep-ex].

\end{thebibliography}

\end{document}